\documentclass[conference]{IEEEtran}

\usepackage{graphicx}

\usepackage{subfigure} 

\begin{document}

\title{The Short-term Memory (D.C. Response) of the Memristor Demonstrates the Causes of the Memristor Frequency Effect}


\author{
	\authorblockN
	{
	Ella Gale\authorrefmark{1}\authorrefmark{2},
	Ben de Lacy Costello\authorrefmark{1},
	Victor Erokhin\authorrefmark{3} and
Andrew Adamatzky\authorrefmark{1}
	}
	\authorblockA{
			\authorrefmark{1}Center for Unconventional Computing, University of the West of England, Bristol, UK, BS16 1QY\\
					}
	\authorblockA{\authorrefmark{2}Email:ella.gale@uwe.ac.uk}
			\authorblockA{
			\authorrefmark{3} CNR-IMEM, Parma, Italy}	
}


%

\specialpapernotice{(Invited Paper)}

\maketitle

\begin{abstract}
A memristor is often identified by showing its distinctive pinched hysteresis curve and testing for the effect of frequency. The hysteresis size should relate to frequency and shrink to zero as the frequency approaches infinity. Although mathematically understood, the material causes for this are not well known. The d.c. response of the memristor is a decaying curve with its own timescale. We show via mathematical reasoning that this decaying curve when transformed to a.c. leads to the frequency effect by considering a descretized curve. We then demonstrate the validity of this approach with experimental data from two different types of memristors.  
\end{abstract}


%
\IEEEpeerreviewmaketitle

\section{Introduction}
The memristor~\cite{14} is a novel electronic component of interest because it possesses an implicit memory: this has led to the suggestion that it will be useful next-generation computer memory~\cite{15} and of use in mimicking how the brain works. 

The memristor was first defined as a device that directly related charge, $q$ to magnetic flux $\varphi$, and thus is the fourth fundamental circuit element (after the resistor, inductor and capacitor) and the first non-linear one~\cite{14}. From Chua's seminal paper in 1971~\cite{14}, it was theorised that the memristor would give a pinched hysteresis I-V curve (a Lissajous curve) which crosses at the origin. Memristive systems were then defined as a memristor with a second state variable~\cite{84}. Once memristors were knowingly fabricated~\cite{15} and ReRAM explained via memristor theory~\cite{119}, the focus on how to test if a device was a memristor changed. The requirement that the memristor cross at the origin (as opposed to being merely pinched at the origin) was relaxed~\cite{8,276} with suggestions that it might not be necessary condition to define a memristor~\cite{291} and that real devices can posses a `nanobattery'~\cite{292}. The frequency test was expanded: 2 of the 3 fingerprints suggested to define the memristor~\cite{276} are to do with frequency behaviour. These are that above some critical frequency, $\omega^{*}$, the hysteresis 
lobe area should decrease monotonically and the pinched hysteresis loop should shrink to zero when the frequency tends to infinity~\cite{276}. Chua suggested that researchers should plot $q$ and $\varphi$ to prove that they have a memristor, however, due to the difficulty in understanding which magnetic flux was relevant~\cite{F1} and the fact that it's more standard to plot voltage-current curves, few people currently do. However, Chua then suggested that to prove if a device was a memristor, the frequency effect should be tested for -- this idea has gained acceptance amongst memristor experimentalists. 

\begin{figure}
\centerline{\subfigure[PEO-PANI 3-terminal memristive system]{\includegraphics[width=1.65in]{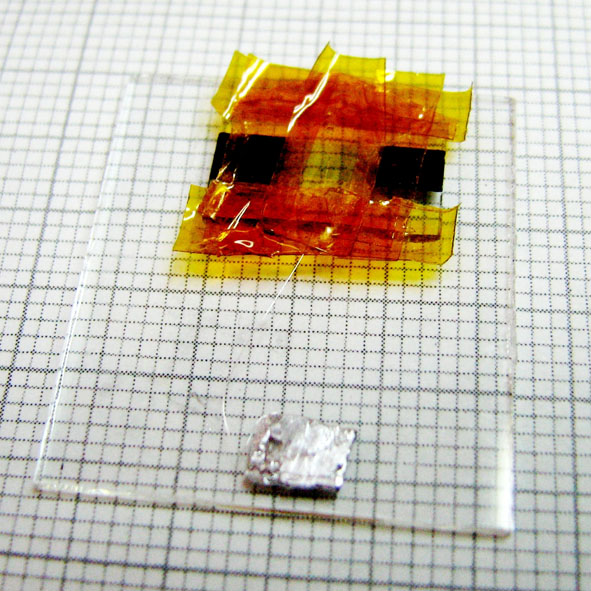}
\label{fig_first_case}}
\hfil
\subfigure[TiO$_2$ sol-gel memristors]{\includegraphics[width=1.65in]{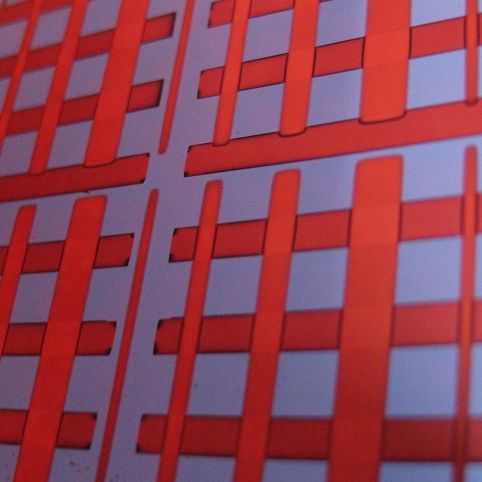}
\label{fig_second_case}}}
\caption{Devices Tested.}
\label{fig:Devices}
\end{figure}

The frequency effect has been derived from memristor theory, however the physics of the effect is not well-understood and have not been related to the memristor material properties. The memristor is commonly thought to be an a.c. component, however we have been experimentally investigating the d.c. response of the memristor~\cite{SpcJ}, which is the route of the short-term memory of the memristor. In this short paper, we shall show how the hysteresis is related to the d.c. response of the memristor (its short-term memory), discuss how the material properties of the device give rise to this effect and finally demonstrate the effect using experimental data from real devices.

The data presented in this paper come from two different types of memristors, see figure~\ref{fig:Devices}: the flexible TiO$_2$ sol-gel memristors described~\cite{260} and the plastic electronic PEO-PANI memristors~\cite{12,23,51} (technically a memristive system~\cite{84} under strict definitions). Note that, the TiO$_2$ sol-gel memristors appear similar to a non-faradic capacitor~\cite{293} over this range, but that standard seemingly `pinched' (the devices actually approach zero but do not cross at zero due to either lag in the ionic current, presence of a nanobattery arising from the electrochemistry~\cite{292}) memristor curve is observed over larger ranges~\cite{260}.

\section{Methodology}

\begin{figure}
\centering
\includegraphics[width=3.2in]{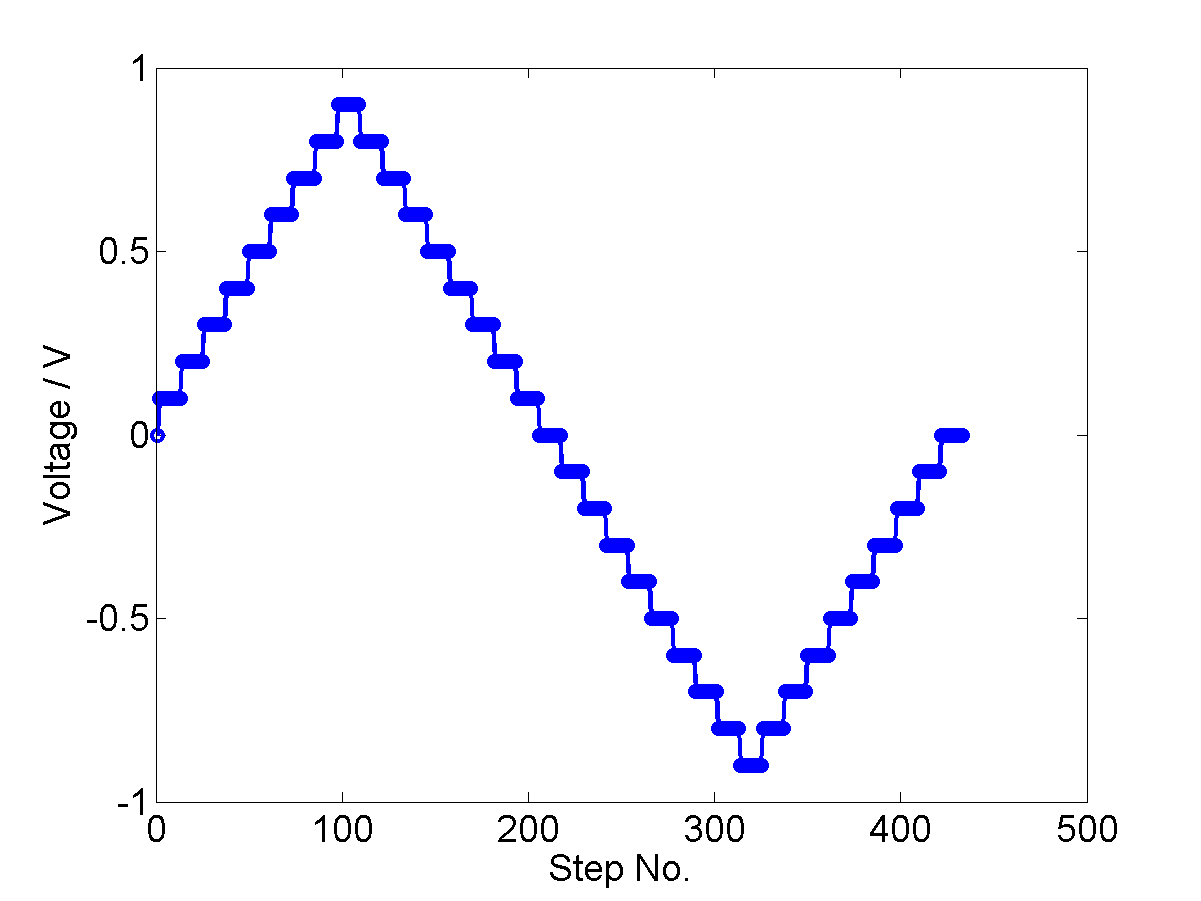}
\caption{Voltage waveform used for PEO-PANI measurements.}
\label{fig:Vt}
\end{figure}

The dynamics of the two memristors are different, so have been measured with different values. In both cases, the voltage waveform was a triangular waveform with $x$ measurement steps, $\Delta t$, at each voltage step $\Delta v$, an example $V-t$ waveform is shown for PEO-PANI in figure~\ref{fig:Vt}. The PEO-PANI memristors were measured on a Keithley programmable multimeter, the total number of measurement steps, $N$ was $N=432$, $x = 12$, $\Delta t=20s$, $\Delta V$=0.1V, with a maximum voltage $\pm 0.9V$. The TiO$_2$ memristors were measured using a Keithley 2400 sourcemeter, N=1600, $\Delta t=0.01s$, $x=10$, $\Delta V=0.0375$, with a maximum of $\pm V 1.5V$. The Keithley was used in auto-zero mode for maximum accuracy, which adds a 0.6s padding time during which the sourcemeter auto-zeroed before taking the reading and the time step was then applied after that. The machine was operated such that the current range was set so there was no extra time searching for the correct current range (which can happen in some set-ups).

\section{Mathematical Description of the System}

\begin{figure}
\centering
\includegraphics[width=3.2in]{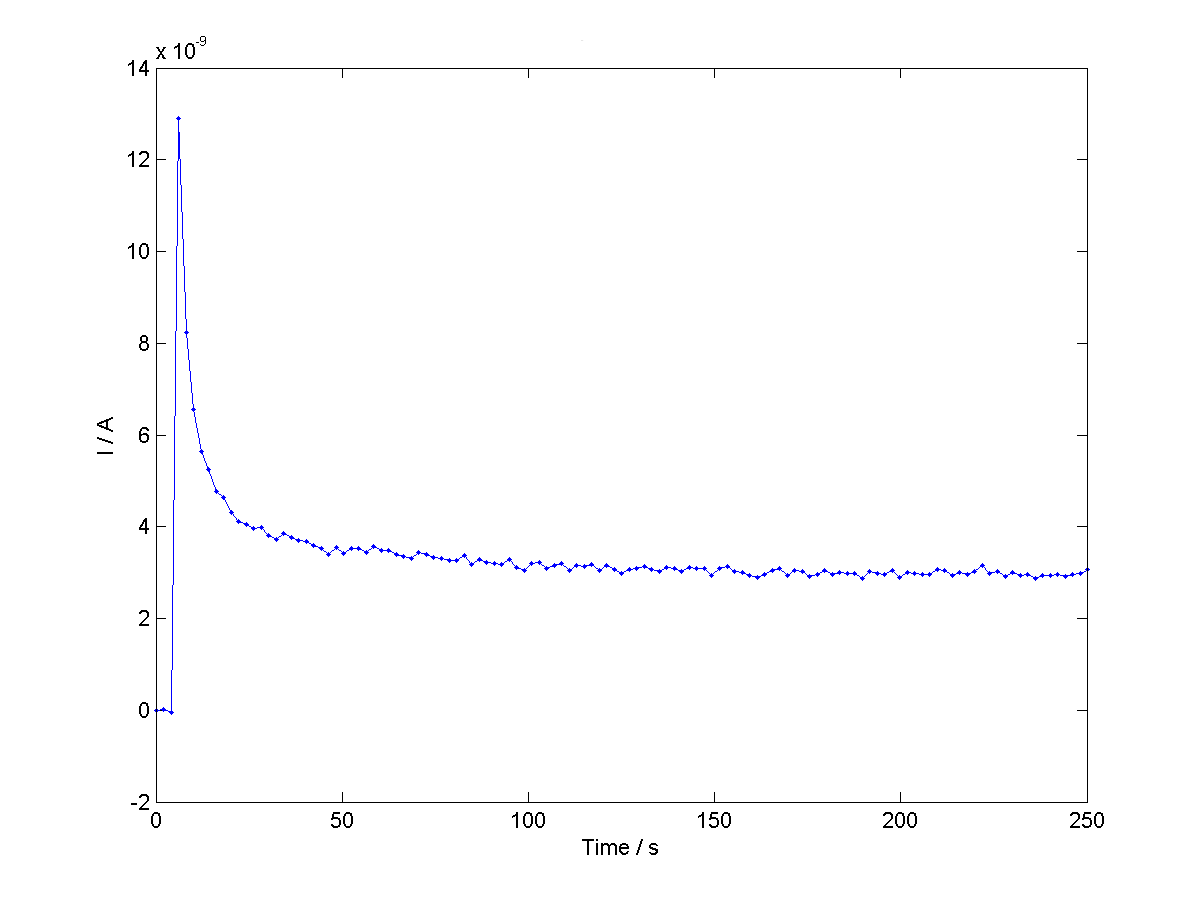}
\caption{d.c. response of the memristor. The device was subjected to a voltage step from 0V to +0.5V after 4 time-steps, $\Delta t=0.05s$ in this measurement.}
\label{fig:DC}
\end{figure}

For each voltage step, $\Delta v$, the voltage is first changed at $t=0$, then the current response is measured at $i_1$, (which would be at $t=0$ if the auto-zero time were turned off), then further measurements made an integer number of time-steps after at $i_2$, $i_3$ ... $i_x$ where $x$ is the number of measurement points at each voltage step. As the size of $\Delta v \rightarrow \delta v$, the current response tends to the continuum (a.c.).

The d.c. response of the memristor (for the TiO$_2$ sol-gel device) is shown in figure~\ref{fig:DC} and has been discussed at length in~\cite{SpcJ}. The measured current is the response to a voltage step and the features are: $i_{\mathrm{max}}$, the maximum size of the peak; $\tau_{\infty}$, the time after which the current is at its equilibrium value (i.e. the value at $t=\infty$ if nothing else changes); the set of measurements we make which is $\{ i(\Delta t) \}$; and the continuous time ($t$) decay dynamics given by $I(t)$. We have shown that $i(\Delta t)$ can be fit by a memristor model~\cite{254}, but in this work we assume that we do not know the form of $I(t)$. The d.c. response of the memristor, can also be described as its short-term memory, and $\tau$ is the persistence time of that memory.

\begin{figure}
\centering
\includegraphics[width=3.2in]{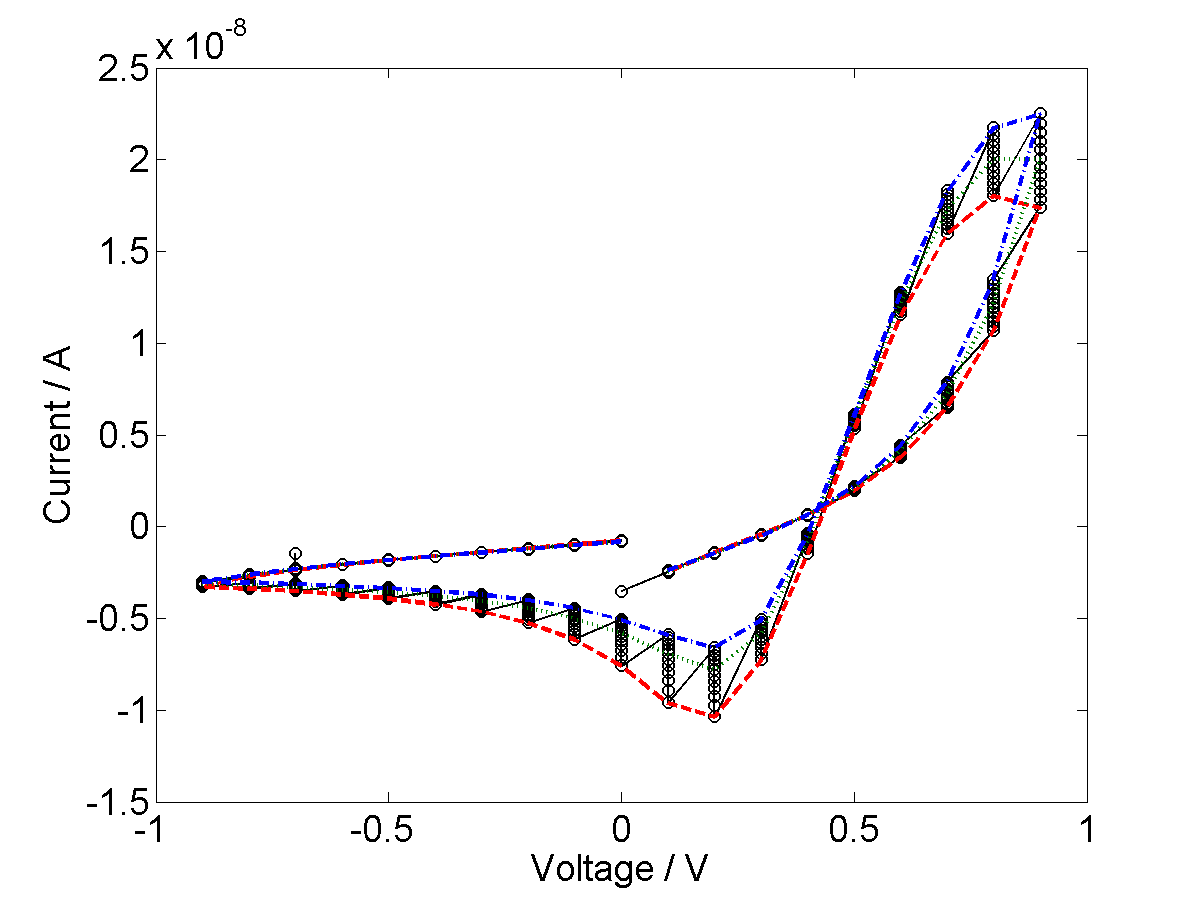}
\caption{PEO-PANI Memristor No. 1. Key:- Red dashes: $i(t_1)$; Green dots: $i(t_6)$; blue dot-dashes: $i(t_{12})$}
\label{fig:PP_1}
\end{figure}

\begin{figure}
\centering
\includegraphics[width=3.2in]{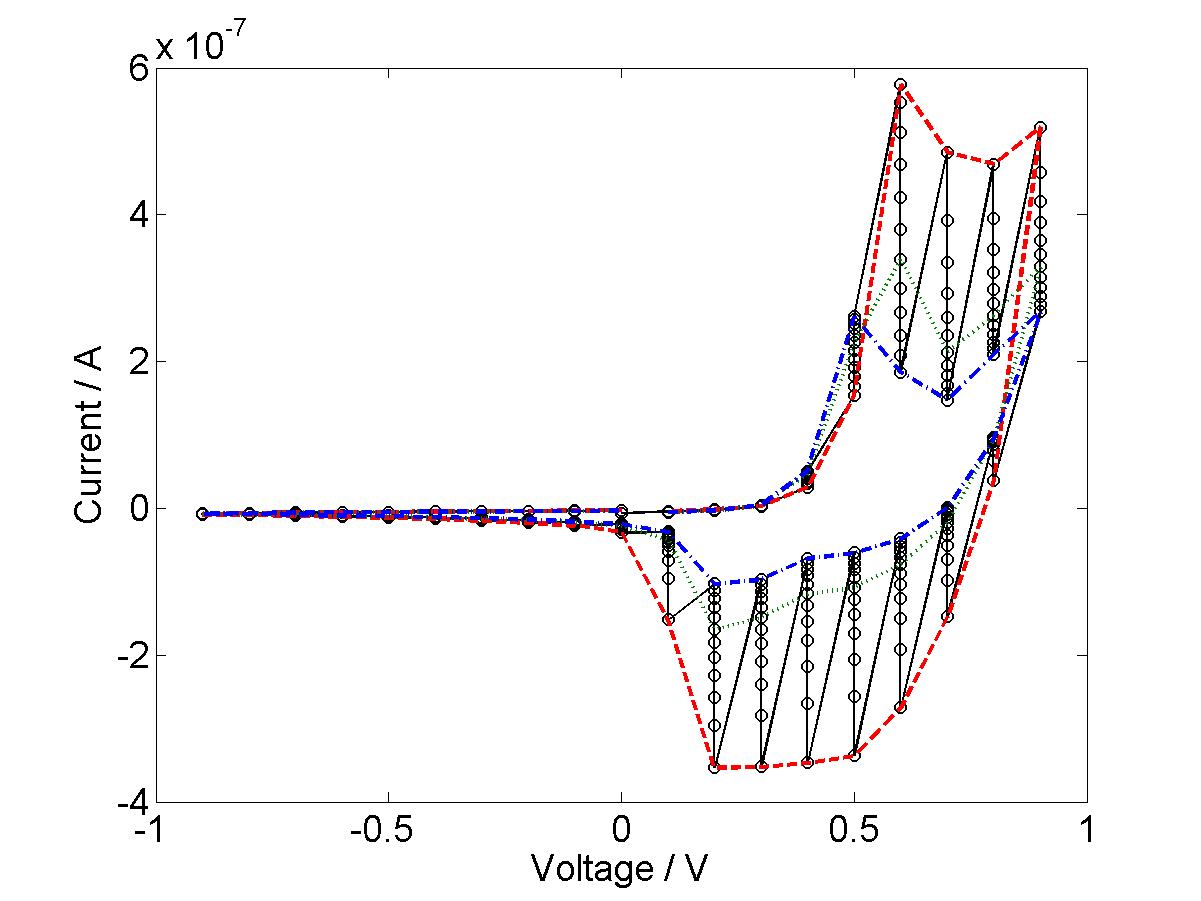}
\caption{PEO-PANI Memristor No. 2 Key:- Red dashes: $i(t_1)$; Green dots: $i(t_6)$; blue dot-dashes: $i(t_{12})$}
\label{fig:PP_2}
\end{figure}

\section{Experimental data}

\begin{figure}
\centering
\includegraphics[width=3.2in]{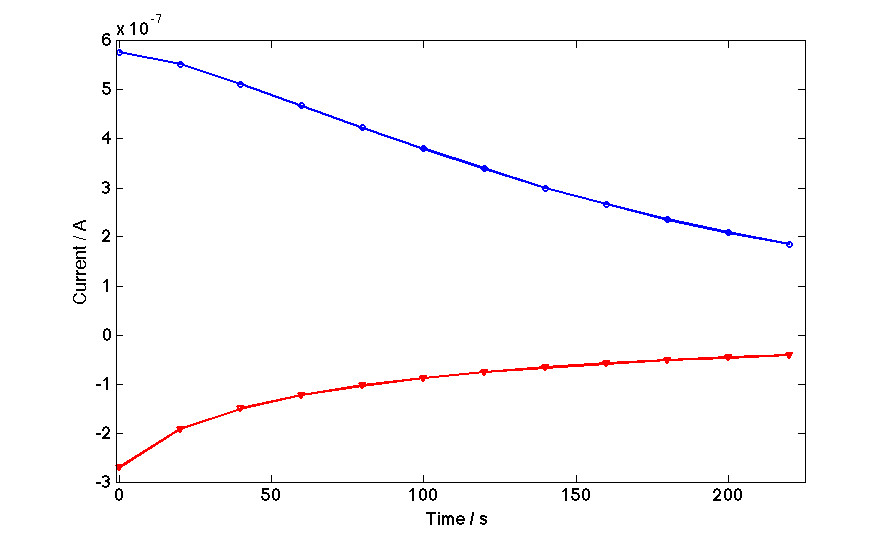}
\caption{`Decay' dynamics at $\pm$0.6V on the PEO-PANI memristor No. 2. Blue: response to +0.6V, Red: response to -0.6V. Similar decay curves are seen in all PEO-PANI devices measured.}
\label{fig:PP_Slice}
\end{figure}

Figures~\ref{fig:PP_1} and ~\ref{fig:PP_2} show the response of two different PEO-PANI devices. The first ($t=1$), middle ($t=6$) and last ($t=12$) measurement points at each voltage have been joined together with red, green and blue lines. At each voltage step, as time passes, the hysteresis associated with it decreases (although it seems to shift in the positive lobe of PEO-PANI memristor 1, rather than decrease). An $I-t$ slice through this I-V curve at V=+0.6 is shown in figure~\ref{fig:PP_Slice}: these dynamics are consistent with those seen under steady state voltage  experiments with these devices.

\begin{figure}
\centering
\includegraphics[width=3.2in]{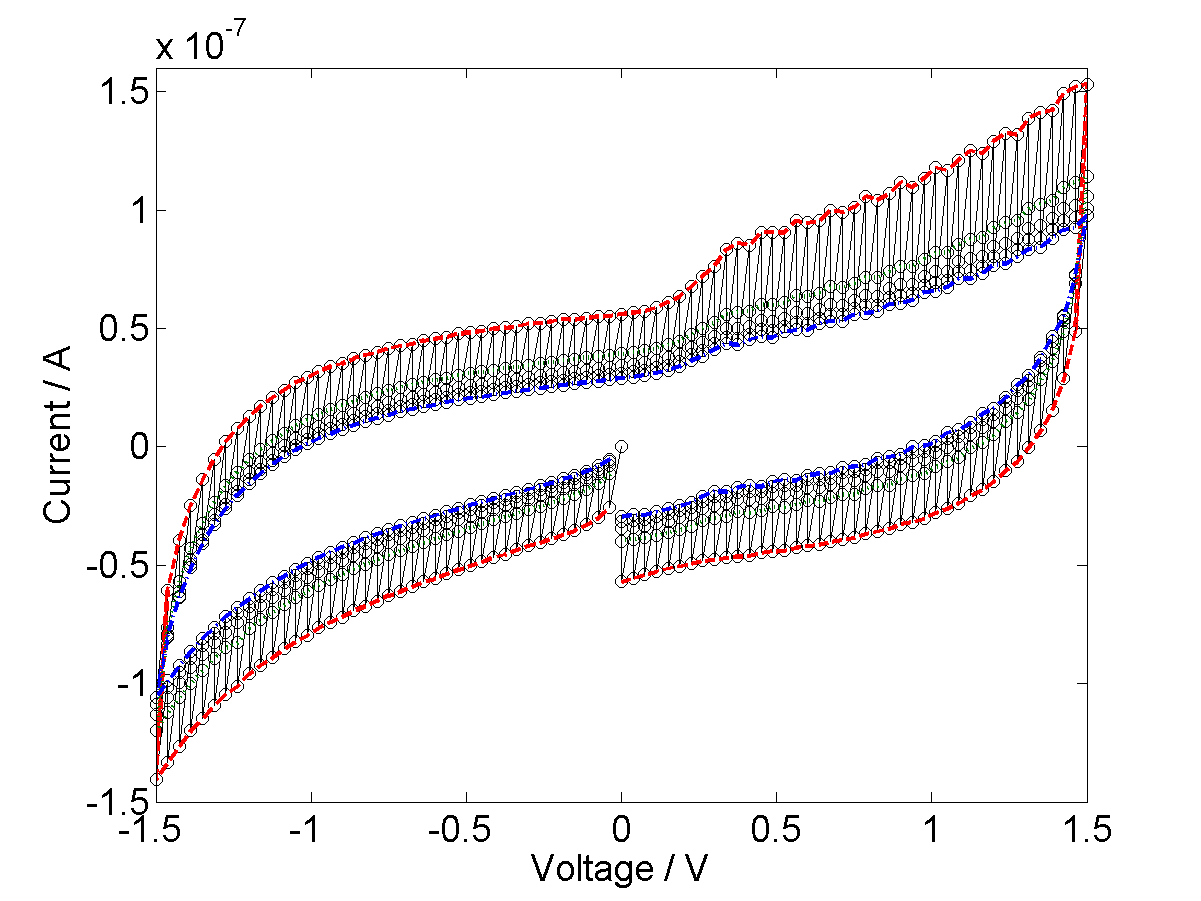}
\caption{TiO$_2$ sol-gel memristor I-V curves. Key:- Red dashes: $i(t_1)$; Green dots: $i(t_4)$; blue dot-dashes: $i(t_10)$}
\label{fig:TiO2}
\end{figure}

Figure~\ref{fig:TiO2} shows the response for a flexible TiO$_2$ device that exhibits open loop behaviour. The dynamics are faster than PEO-PANI, but the behaviour is qualitatively the same: as dwell time at each voltage increases the magnitude of the current decreases, leading to a shrinking hysteresis.

\section{Discussion}

These data are for step $V-t$ waveforms (the experimental reason for this is that it is impossible to source slow enough alternating voltage waveforms), but if we move from the descretized time measurement domain to a continuous time domain, we can investigate the frequency response. 

For our set of, for example, 12 measurement points, we have the following data points for $i$: $\{ i(\Delta t), i(2 \Delta t), ... i(12\Delta t) \}$. By drawing a line between points with the same value of $x$ (where $x$ is the measurement point) we get the following set of frequencies: $\{ \omega \} = \{ (x N \Delta t)^{-1} \}$, where $x = 1,2... 12$. In the PEO-PANI example, these frequencies are given by: $\{ \omega \} = \{ 1, \frac{1}{2}, \frac{1}{3}, \frac{1}{4} ... \frac{1}{12} \} $, in units of $\frac{1}{N \Delta t}$; which is actually a set of slower (larger) frequencies. So, by plotting the lines between adjacent measurements with the same $x$ value, we are actually approximating different frequency responses. And, we see from the data that as the frequency gets larger the hysteresis, $H$, decreases. 

The frequency response calculated in this way is approximate however, because we are actually getting a measure of the state of the memristor (or the amount of short-term memory it has) after that time, we are not including the response to a changing voltage. Specifically, at voltage step $n$ measurement step $x$ we get the response to the voltage input at $t=0$ (which was $x$ time-steps before). For an actual measurement of the corresponding frequency we would have the response to both the voltage input at $t=0$ and the voltage input at that measurement step ($x$) and that frequency. Thus we need to know the effect of a changing voltage on this memory and there are two ways to do this. The first is to actually measure the memristor at several different frequencies, which we have done (not shown here for space reasons) and which shows that slower frequencies tend to decrease the hysteresis as expected. The second is to take account of the interaction between a voltage input and the memristor's short-term memory, which we know from previous work~\cite{P0c} means that the response is smaller than it would be if there was no memory, this effect does not change the qualitative results shown here, and so, acknowledging these facts, we can still proceed to see what further information these plots can give us.

As $\Delta t \rightarrow \delta t$ (i.e. we go from descretized time to continuous time), $v(\Delta t)$ becomes a triangular waveform and $\{ i(\Delta t \}$ becomes $I(t)$ and spikes are smoothed into a curve, similar to those drawn by joining points with the same value of $x$ and $\{ \omega \}$ becomes a continuum. Thre result that the hysteresis decreases as the frequency gets larger, becomes $\omega \rightarrow \infty, H \rightarrow 0$, which is the third fingerprint of the memristor. The observations that this effect exists with a descretized waveform, that the points have similar dynamics to those seen in $I-t$ plots and match results for different frequency tests illustrates that the short-term memory (d.c. response) of the memristor is relevant for understanding the frequency effect. 

\subsection{What is the Single Valued Function?}

The third fingerprint of the memristor is that the `pinched hysteresis loop should shrink to a single-valued function when the frequency tends to infinity'~\cite{276}, what is this single valued function?  At slow frequencies, the memory of the memristor is lost at each `time-step' and the measured current is that of $i(\tau_{\infty})$. As $\omega \rightarrow \infty$, $\{ i \Delta t \} \rightarrow \{i(\tau_{\infty},v_1), i(\tau_{\infty},v_2), ... i(\tau_{\infty},v_N) \}$, and because $i(\tau_{\infty},v_k) =i(\tau_{\infty},v_l)$, if $v_k=v_l$ there is no hysteresis. This is because the different current values seen at $v_k$ and $v_l$ (if say $v_k$ is +0.6V on the first quadrant of the $V-t$ plot and $v_l$ is +0.6V on the second, i.e. after the device has been taken to $V_{\mathrm{max}}$ and back) are entirely a result of the memory in the system which is stored as the short-term memory and essentially different actions applied to the device move the current response up and down the $I-t$ curve shown in figure~\ref{fig:DC}. Thus, moving from descretized time to continuous time as before, we get $\omega \rightarrow \infty, H \rightarrow 0,$ and $I(t) \leftarrow \{ i \Delta t \} \rightarrow \{i(\tau_{\infty},v_1), i(\tau_{\infty},v_2), ... i(\tau_{\infty},v_N) \}$, or the memory-less response, which is a straight-line in $I-V$ space corresponding to an ideal ohmic resistor with resistance of the equilibrated resistor, $M(\tau_{\infty})$. 

\subsection{What is the Maximal Value of the Hysteresis?}

From examination of the $I-t$ plot we can answer the question of what happens in the short-time limit. At the shortest time we can measure (around 0.05s in the memristor devices using the current set-up) we see the maximum spike current, $i_{\mathrm{max}}$, and if we measure at the corresponding frequency, we will see the largest value of the hysteresis. We postulate that there is a resonant frequency, $\omega_0$, where we will get the largest hysteresis, and this corresponds to the systems being as close to the peak of $i_{\mathrm{max}}$ in d.c. space as possible. Below this frequency, we see the hysteresis decreasing with frequency (as shown here experimentally). If this $\omega_0$ is the same as the critical frequency $\omega^{*}$, then this would match the limits of the second fingerprint of the memristor. There are reasons to believe that $\omega_0$ is not close to 0 and that the hysteresis decreases above it as well as below it, specifically, in that it takes time for the memory property of the system 
to respond.

\subsection{Material Causes of the Frequency Effect}

Memristance arises due to the slower response time of the memory-carrying ions/vacancies than the conducting electrons to a voltage change. Thus memristance can be described as a two level system involving the interaction of two (or more) charge carrying particles with different mobilities. For TiO$_2$ memristors we believe that the timescale of memristance is related to the ion mobility of oxygen vacancies (a point of view also held by others~\cite{15}). The oxygen ions take time to respond to a voltage change, due to their mass. Thus, if an applied frequency is so fast that the period is less than the time taken for the oxygen ions to change direction, no appreciable change in resistance will be seen and the hysteresis will shrink to 0. The maximal value of the hysteresis would thus be seen when the frequency was able to move the oxygen ions the furthest over a period. Straight-line resistor-like $I-V$ responses have been recorded by us when the frequency is too fast for the system (not shown in this short paper). For PEO-PANI memristors a similar argument holds, but the memory carrying ions are positive ions, not oxygen vacancies.

\section{Conclusion}

This type of I-V curve offers information about the dynamics of the system and demonstrates the frequency effect, by virtue of being a single graph from a single experiment and we suggest including this type of plot in memristor papers generally. We highlight that the d.c. response of the memristor gives information about its short-term memory and it is the interaction of this memory with the frequency of measurement (which is related to the frequency of the $V-t$ curve in the continuum case) which causes the frequency effect. Finally, we suggest (and discuss elsewhere~\cite{F1}) that this frequency effect arises different timescales for ions and electrons.

\bibliographystyle{IEEEtran.bst}

\maketitle
\bibliography{IEEEabrv,./UWELit}

\end{document}